\documentstyle[twoside,epsf]{article}

\catcode`\@=11
\long\def\@makefntext#1{
\protect\noindent \hbox to 3.2pt {\hskip-.9pt  
$^{{\eightrm\@thefnmark}}$\hfil}#1\hfill}		

\def\@makefnmark{\hbox to 0pt{$^{\@thefnmark}$\hss}}	
	
\def\ps@myheadings{\let\@mkboth\@gobbletwo
\def\@oddhead{\hbox{}
\rightmark\hfil\eightrm\thepage}   
\def\@oddfoot{}\def\@evenhead{\eightrm\thepage\hfil
\leftmark\hbox{}}\def\@evenfoot{}
\def\sectionmark##1{}\def\subsectionmark##1{}}

\oddsidemargin=\evensidemargin
\newcounter{sectionc}\newcounter{subsectionc}\newcounter{subsubsectionc}
\renewcommand{\section}[1] {\vspace{12pt}\addtocounter{sectionc}{1} 
\setcounter{subsectionc}{0}\setcounter{subsubsectionc}{0}\noindent 
	{\tenbf\thesectionc. #1}\par\vspace{5pt}}
\renewcommand{\subsection}[1] {\vspace{12pt}\addtocounter{subsectionc}{1} 
	\setcounter{subsubsectionc}{0}\noindent 
	{\bf\thesectionc.\thesubsectionc. {\kern1pt \bfit #1}}\par\vspace{5pt}}
\renewcommand{\subsubsection}[1] {\vspace{12pt}\addtocounter{subsubsectionc}{1}
	\noindent{\tenrm\thesectionc.\thesubsectionc.\thesubsubsectionc.
	{\kern1pt \tenit #1}}\par\vspace{5pt}}
\newcommand{\nonumsection}[1] {\vspace{12pt}\noindent{\tenbf #1}
	\par\vspace{5pt}}

\newcounter{appendixc}
\newcounter{subappendixc}[appendixc]
\newcounter{subsubappendixc}[subappendixc]
\renewcommand{\thesubappendixc}{\Alph{appendixc}.\arabic{subappendixc}}
\renewcommand{\thesubsubappendixc}
	{\Alph{appendixc}.\arabic{subappendixc}.\arabic{subsubappendixc}}

\renewcommand{\appendix}[1] {\vspace{12pt}
        \refstepcounter{appendixc}
        \setcounter{figure}{0}
        \setcounter{table}{0}
        \setcounter{lemma}{0}
        \setcounter{theorem}{0}
        \setcounter{corollary}{0}
        \setcounter{definition}{0}
        \setcounter{equation}{0}
        \renewcommand{\thefigure}{\Alph{appendixc}.\arabic{figure}}
        \renewcommand{\thetable}{\Alph{appendixc}.\arabic{table}}
        \renewcommand{\theappendixc}{\Alph{appendixc}}
        \renewcommand{\thelemma}{\Alph{appendixc}.\arabic{lemma}}
        \renewcommand{\thetheorem}{\Alph{appendixc}.\arabic{theorem}}
        \renewcommand{\thedefinition}{\Alph{appendixc}.\arabic{definition}}
        \renewcommand{\thecorollary}{\Alph{appendixc}.\arabic{corollary}}
        \renewcommand{\theequation}{\Alph{appendixc}.\arabic{equation}}
        \noindent{\tenbf Appendix \theappendixc #1}\par\vspace{5pt}}
\newcommand{\subappendix}[1] {\vspace{12pt}
        \refstepcounter{subappendixc}
        \noindent{\bf Appendix \thesubappendixc. {\kern1pt \bfit #1}}
	\par\vspace{5pt}}
\newcommand{\subsubappendix}[1] {\vspace{12pt}
        \refstepcounter{subsubappendixc}
        \noindent{\rm Appendix \thesubsubappendixc. {\kern1pt \tenit #1}}
	\par\vspace{5pt}}

\newcommand{\textlineskip}{\baselineskip=13pt}
\newcommand{\smalllineskip}{\baselineskip=10pt}

\def\eightcirc{
\begin{picture}(0,0)
\put(4.4,1.8){\circle{6.5}}
\end{picture}}
\def\eightcopyright{\eightcirc\kern2.7pt\hbox{\eightrm c}} 

\newcommand{\copyrightheading}[1]
	{\vspace*{-2.5cm}\smalllineskip{\flushleft
	{\footnotesize International Journal of Modern Physics B, #1}\\
	{\footnotesize $\eightcopyright$\, World Scientific Publishing
	 Company}\\
	 }}

\newcommand{\publisher}[2]{{\begin{center}\footnotesize\smalllineskip 
	Received #1\\
	Revised #2
	\end{center}
	}}

\def\abstracts#1#2#3{{
	\centering{\begin{minipage}{4.5in}\baselineskip=10pt\footnotesize
	\parindent=0pt #1\par 
	\parindent=15pt #2\par
	\parindent=15pt #3
	\end{minipage}}\par}} 

\renewenvironment{thebibliography}[1]			
	{\frenchspacing
	 \ninerm\baselineskip=11pt
	 \begin{list}{\arabic{enumi}.}
	{\usecounter{enumi}\setlength{\parsep}{0pt}
	 \setlength{\leftmargin 12.7pt}{\rightmargin 0pt} 
	 \setlength{\itemsep}{0pt} \settowidth
	{\labelwidth}{#1.}\sloppy}}{\end{list}}

\newcounter{itemlistc}
\newcounter{romanlistc}
\newcounter{alphlistc}
\newcounter{arabiclistc}

\newcommand{\fcaption}[1]{
        \refstepcounter{figure}
        \setbox\@tempboxa = \hbox{\footnotesize Fig.~\thefigure. #1}
        \ifdim \wd\@tempboxa > 5in
           {\begin{center}
        \parbox{5in}{\footnotesize\smalllineskip Fig.~\thefigure. #1}
            \end{center}}
        \else
             {\begin{center}
             {\footnotesize Fig.~\thefigure. #1}
              \end{center}}
        \fi}

\newcommand{\tcaption}[1]{
        \refstepcounter{table}
        \setbox\@tempboxa = \hbox{\footnotesize Table~\thetable. #1}
        \ifdim \wd\@tempboxa > 5in
           {\begin{center}
        \parbox{5in}{\footnotesize\smalllineskip Table~\thetable. #1}
            \end{center}}
        \else
             {\begin{center}
             {\footnotesize Table~\thetable. #1}
              \end{center}}
        \fi}

\def\@citex[#1]#2{\if@filesw\immediate\write\@auxout
	{\string\citation{#2}}\fi
\def\@citea{}\@cite{\@for\@citeb:=#2\do
	{\@citea\def\@citea{,}\@ifundefined
	{b@\@citeb}{{\bf ?}\@warning
	{Citation `\@citeb' on page \thepage \space undefined}}
	{\csname b@\@citeb\endcsname}}}{#1}}

\newif\if@cghi
\def\cite{\@cghitrue\@ifnextchar [{\@tempswatrue
	\@citex}{\@tempswafalse\@citex[]}}
\def\citelow{\@cghifalse\@ifnextchar [{\@tempswatrue
	\@citex}{\@tempswafalse\@citex[]}}
\def\@cite#1#2{{$\null^{#1}$\if@tempswa\typeout
	{IJCGA warning: optional citation argument 
	ignored: `#2'} \fi}}

\def\pmb#1{\setbox0=\hbox{#1}
	\kern-.025em\copy0\kern-\wd0
	\kern.05em\copy0\kern-\wd0
	\kern-.025em\raise.0433em\box0}

\def\fnt#1#2{\footnotetext{\kern-.3em
	{$^{\mbox{\scriptsize #1}}$}{#2}}}

\def\fpage#1{\begingroup
\voffset=.3in
\thispagestyle{empty}\begin{table}[b]\centerline{\footnotesize #1}
	\end{table}\endgroup}

\def\runninghead#1#2{\pagestyle{myheadings}
\markboth{{\protect\footnotesize\it{\quad #1}}\hfill}
{\hfill{\protect\footnotesize\it{#2\quad}}}}
\font\tenrm=cmr10
\font\tenit=cmti10 
\font\tenbf=cmbx10
\font\bfit=cmbxti10 at 10pt
\font\ninerm=cmr9

\font\eightrm=cmr8

\def\qed{\hbox{${\vcenter{\vbox{			
   \hrule height 0.4pt\hbox{\vrule width 0.4pt height 6pt
   \kern5pt\vrule width 0.4pt}\hrule height 0.4pt}}}$}}

\def\bsc{{\sc a\kern-6.4pt\sc a\kern-6.4pt\sc a}}	
\def\bflatex{\bf L\kern-.30em\raise.3ex\hbox{\bsc}\kern-.14em 
T\kern-.1667em\lower.7ex\hbox{E}\kern-.125em X} 

\begin{document}

\runninghead{Polaron Crystallization and Melting}
{Polaron Crystallization and Melting: Effects of the Long-Range Coulomb 
Forces}

\normalsize\textlineskip
\thispagestyle{empty}
\setcounter{page}{1}

\copyrightheading{}			

\vspace*{0.88truein}

\fpage{1}
\centerline{\bf POLARON CRYSTALLIZATION AND MELTING:}
\vspace*{0.035truein}
\centerline{\bf EFFECTS OF THE LONG-RANGE COULOMB FORCES}
\vspace*{0.37truein}
\centerline{\footnotesize S. FRATINI and P. QU\'EMERAIS}
\vspace*{0.015truein}
\centerline{\footnotesize\it Laboratoire d'Etudes des Propri\'et\'es Electroniques
des Solides, CNRS} 
\baselineskip=10pt
\centerline{\footnotesize\it BP 166, 38042 Grenoble Cedex 9, France} 
\vspace*{10pt}
\publisher{30 October 1998}

\vspace*{0.21truein}
\abstracts{On examining the stability of a Wigner crystal in an ionic 
dielectric, two competitive effects due to the polaron formation are found 
to be important: (i) the screening of the Coulomb force $1/\varepsilon_s 
r$, which destabilizes the crystal, compensated by (ii) the increase of the 
carrier mass (polaron mass).  The competition between the two effects is 
carefully studied, and the quantum melting of the polaronic Wigner crystal 
is examined by varying the density at zero temperature.  By calculating the 
quantum fluctuations of both the electron and the polarization, we show 
that there is a competition between the {\it dissociation of the polarons} 
at the insulator-to-metal transition (IMT), and a melting towards a polaron 
liquid.  We find that {\it at strong coupling, a liquid state of dielectric 
polarons cannot exist}, and the IMT is driven by the polaron dissociation.  
Next, taking into account the dipolar interactions between localized 
carriers, we show that these are responsible for an instability of the 
transverse vibrational modes of the polaronic Wigner crystal as the density 
increases.  This provides a new mechanism for the IMT in doped dielectrics, 
which yields interesting dielectric properties below and beyond the 
transition.  An optical signature of such a mechanism for the IMT is 
provided.
}{}{}

\vspace*{10pt}


\vspace*{1pt}
\textlineskip	
\section{Introduction}
\vspace*{-0.5pt}
\noindent

In a recent letter to Nature, Phillips {\it et al.}\cite{phillips} pointed 
out that a superconducting state could be expected close to the instability 
of a 2D Wigner Crystal (WC) of electrons.  Previously, Takada\cite{takada} 
had mentioned such a possibility for a 3D WC.  In both cases, the pairing 
mechanism is due to correlation between electrons, mediated by the long 
range Coulomb interaction.  More explicitely, pairing occurs through the 
Coulomb holes which form in the vicinity of the melting transition.  
Translated into the language of dielectric constants, as was first 
recognized by Bagchi\cite{bagchi} for the 3D WC, it means that the static 
dielectric constant is negative.  In his original paper, Bagchi also 
extended his results to an insulating lattice of Lorentz dipoles.  He found 
that, on increasing the dipole density, the Lorentz lattice undergoes a 
phonon instability due to the dipolar interactions.  At the same time, the 
static dielectric constant becomes identical to the dielectric constant of 
an ordinary WC, and thus can be negative.  In other words, an ordinary WC 
appears to be nothing but a Lorentz lattice of dipoles just on the verge of 
instability.  This could provide a natural pairing mechanism for 
(hypothetical) free electrons moving through such a Lorentz lattice close 
to its critical point.  Quite generally, the question of the sign of the 
static dielectric constants has been addressed by many authors.  Cohen and 
Anderson\cite{cohen} argued on physical grounds that such a negative sign 
would be impossible.  Later, the "russian school", in particular Dolgov 
{\it et al.}\cite{dolgov}, following the previous work of 
Martin\cite{martin}, pointed out that a negative sign of the dielectric 
constant would not contradict any stability criterion.  The only required 
condition is that $\varepsilon \left( 0,0 \right) $ be strictly positive: a 
system can have $\varepsilon \left({\bf k},0 \right) <0$ for any $k>0$, and 
still be stable.

We have developed\cite{quem} such a physical picture in order to describe 
the Insulator-to-Metal transition in ionic dielectrics, taking into account 
the formation of large polarons when introducing the doping charges.  These 
polarons, which are ordered in a crystallized state at low densities, 
behave more or less as Lorentz dipoles, and can yield some of the 
interesting physical properties already mentioned by Bagchi.  In order to 
understand our point of view, it is useful to connect our studies to the 
recent "fight" in Physical Review Letters of Chakraverty {\it et al.} 
against the (bi)polaronic superconductivity scenario\cite{alexandrov}.  
These authors did not find the bipolaronic scenario suitable for the 
interpretation of the experimental data in high-Tc, a point of view which 
seems to be fully shared by Anderson\cite{anderson}.  In fact, we would 
like to underline that the basic building block of the (bi)polaronic 
scenario is the existence of a quantum liquid of (bi)polarons.  As we will 
explain later, this turns out to be impossible at strong electron-phonon 
coupling, owing to the drastic role of long-range Coulomb forces in the 
many-polaron problem.  Indeed, treating self-consistently a finite density 
of {\it interacting} polarons is a much more complicated problem than 
considering only one or two polarons.  Extensions to finite densities 
usually neglect the Coulomb interaction between (bi)polarons, as in the 
framework of the random phase approximation\cite{alexandrov}.  As we will 
see, this approximation is quite dubious at strong polar electron-phonon 
coupling, while it should remain reasonable at small and intermediate 
coupling\cite{iadonisi}.  Long-range Coulomb forces are responsible for 
{\it both} the polaron formation (if one only considers the coupling of the 
electrons to the polar mode of the crystal, as it is the case in all our 
studies), and for their mutual interactions.  In such systems, 
electron-phonon and electron-electron interactions cannot be decoupled and 
must be treated on the same ground, otherwise the collective behavior of 
polarons at finite density can be completely missed.  To tackle the 
many-polaron problem, we have considered a system at low density.  Starting 
from the remark\cite{quem,remova} that two polarons at a distance $d$ repel 
as $1/ {\varepsilon_s} d$ if the dielectric constants satisfy 
$\varepsilon_{\infty}/\varepsilon_s >\approx 0.1$ (as it is 
well-known\cite{bipol}, there is no bipolaron formation in that case), {\it 
at low densities, the crystallized state is necessarily the ground-state}.

Let us consider a WC, consisting of a system of electrons plus a jellium 
(assumed to be rigid) of positive charge with the same density.  We can 
imagine that this WC is inserted into a host polar material, characterized 
by the two limiting dielectric constants $\varepsilon_s$, 
$\varepsilon_{\infty}$ and by the frequency of the longitudinal optical 
mode $\omega_{LO}$.  If one turns on the coupling between the WC and the 
polar phonon mode of the host material, two competitive effects arise.  The 
first one corresponds to the screening of the Coulomb interaction between 
the electrons, which is reduced from $1/\varepsilon_{\infty}r$ to 
$1/\varepsilon_s r$.  This screening is due to the host lattice 
polarization which surrounds the electrons, which is nothing but the 
polaron cloud.  The other important effect is the increase of the carrier 
mass: when the coupling with the lattice is turned on, the original band 
mass $m^*$ becomes $M_P$, the polaron mass, which can be large, especially 
in the strong coupling limit for which the polarons are well-formed.  Now, 
the crystallization is due to the overcoming of the kinetic energy $1 /M_p 
R_s^2$ by the Coulomb repulsion $1/\varepsilon_s R_s$, where $R_s$ is 
related to the density $n$ as usual: $n=3/4\pi R_s^3$.  We see from this 
simple argument that the polaronic crystallized state can be stable up to 
fairly high densities in spite of the screening of the Coulomb forces.  
Furthermore, as given by standard polaron theory, the mass $M_P$ becomes 
infinite as the phonon frequency tends to zero.  In that case, i.e.  when 
$\omega_{LO} \rightarrow 0$, the kinetic energy is always negligible with 
respect to the Coulomb repulsion, and melting towards a liquid state of 
polarons is clearly not possible.

In our studies, we have focused on the quantum melting (at zero 
temperature) of such a polaronic Wigner crystal (PC).  We have found that 
the two competitive effects above give rise to two different scenarios for 
the IMT, which can be easily understood.  When the electron-phonon coupling 
vanishes, we are left with an ordinary WC of electrons.  As the density is 
raised, the quantum fluctuations of the localized particles increase, and 
when they reach a certain magnitude, the crystal melts towards a liquid 
state of electrons.  The effect of the host lattice is negligible in that 
case.  In the opposite limit $\omega_{LO} \rightarrow 0$, the 
electron-phonon coupling becomes infinite.  Its strength is given by the 
usual dimensionless parameter\cite{frohlich} $\alpha =\left({{{m^*} / 
{2\hbar ^3\omega _{LO}}}}\right)^{1/2}{{e^2} / {\tilde \varepsilon }}$, 
where $1 /{\tilde \varepsilon }=1 /\varepsilon_{\infty}-1/{\varepsilon_s}$ 
is the effective dielectric constant responsible for the polaron formation.  
In this limit, each electron is self-trapped in a polarization 
potential-well which is {\it frozen}.  As it is well-known, the latter has 
a {\it coulombic} nature and behaves as $1/{\tilde \varepsilon }r$ at large 
distance.  Since the polarization cannot move, the polarons, i.e.  the 
electron {\it plus} the polarization, cannot go towards a liquid state.  
The IMT is thus driven by the dissocation of the polarons and the screening 
of the polarization potentials by the liberated electrons.  We are in a 
situation which closely resembles the usual Mott transition.  On the basis 
of the same argument as Mott\cite{mott}, the critical density can be 
estimated to be:
\begin{equation}
n_c^{1/3} \left( {\tilde \varepsilon }/{\varepsilon_{\infty}}
\right) R_P \approx 0.25, 
\end{equation}
where $R_P$ is the bound-state radius, which is nothing but the polaron 
radius given at strong coupling by $R_P \approx 3.2 {\tilde \varepsilon} 
{\left( m^*/m_e \right)} a_0$. Here, $a_0$ is the usual Bohr radius 
($0.53 \AA$), and $R_P$ corresponds to a hydrogenic wave-function.

Starting from the limiting cases presented above, it is now important to 
extend the results to any finite electron-phonon coupling (i.e. to any 
finite phonon frequency), and examine 
which mechanism is responsible for the melting of the polaron 
crystal.  At this stage of the theory, we have developed two steps.

\pagebreak

\section{Quantum melting: polaron breaking versus polaron liquid}

First, we have modeled the situation in the PC by considering a single 
polaron localized in a positively charged jellium sphere with density 
$n=3/4 \pi R_s^3$.  In this approach, which is a sort of mean-field 
many-body calculation (see for example the first papers by Wigner on the 
electron crystallization), dipolar interactions and exchange effects are 
neglected.  However, one obtains reliable results on the crystal energy and 
ground state properties, provided the density is low enough (large $R_s$).  
As it was done by Nozieres {\it et al.} for the WC\cite{nozieres}, we have 
applied the Lindemann criterion to this model.  These calculations allow us 
to understand, at least qualitatively, the route that the crystal will 
choose to melt: polaron dissociation versus polaron liquid state.  Of 
course, this method just gives access to the melting scenario, not to the 
ground-state beyond the transition.
%
\begin{figure}[t]
\epsfxsize=11cm 
\centerline{\epsffile{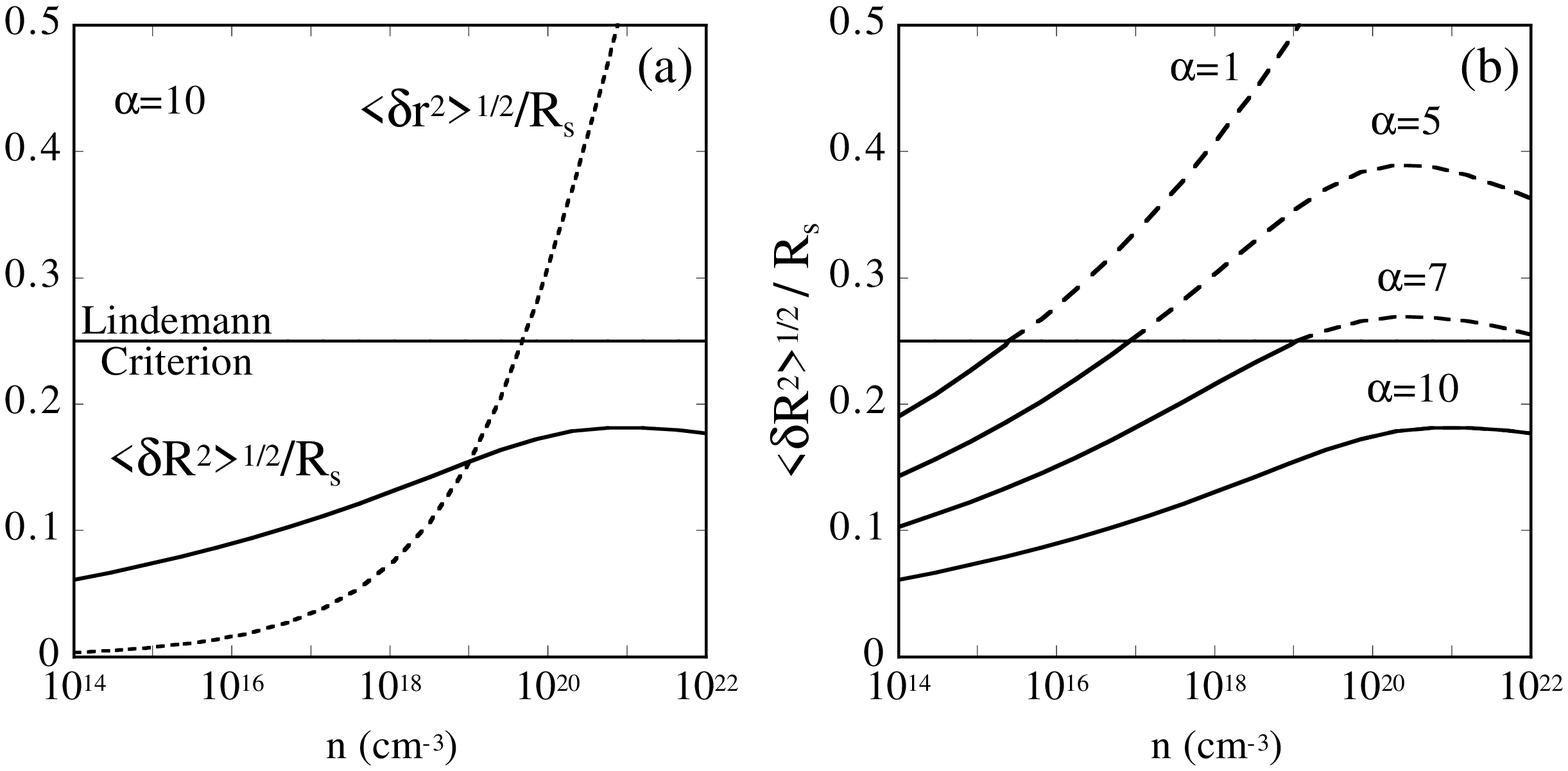}}
\caption{(a)  The fluctuation of the polaron  $\langle \delta R^2
\rangle^{1/2}/R_s$ (bold line) and of its internal degree of freedom $\langle \delta
r^2 \rangle^{1/2}/R_s$ (dotted line) versus $n$ at $\alpha=10$.  (b) The polaron
fluctuation $\langle \delta R^2 \rangle^{1/2}/R_s$ versus $n$ for different $\alpha$.
Physical parameters are $\epsilon_\infty=5$, $\epsilon_s=30$ and 
$m^*=2m_e$. The density parameter is simply $n=1.6\cdot 10^{24}/r_s^3$.}
\label{raggi} 
\end{figure}

The application of this criterion to the PC is more 
complicated than for an ordinary WC, owing to the {\it composite 
nature of the polarons}.  To be able to delocalize the polarons towards a 
liquid state, the quantum fluctuations for each polaron moving as a 
whole, i.e.  the motion of each electron together with its surrounding 
polarization, must become large in comparison with $R_s$.  On the contrary, 
if the relative fluctuations of the electron with respect to the 
polarization become large, the polarons break apart (dissociation).  The 
Feynman treatment for the polaron\cite{feynman}, which relies on path integral 
calculations, yields a natural way to evaluate such quantities: since the 
polarization field in Feynman's approach is replaced by a rigid particle 
with coordinate $X$, one easily calculates both the quantum fluctuations of 
the center of mass $R=(m^*x+MX)/(m^*+M)$ and the relative coordinate 
$r=x-X$ ($x$ is the electron coordinate).  Thus, we now have two different 
Lindemann criteria, which correspond to two different melting scenarios:

(i) ${\left < \delta R^2 \right >}^{1/2}/R_s>1/4$ which allows the melting of
the crystal towards a liquid state of polarons;

(ii) ${\left < \delta r^2 \right>}^{1/2}/R_s>1/4$ for the polaron
dissociation.

Both quantities are calculated by the same path integral variational 
procedure\cite{quem,feynman}.

The frequencies of the two degrees of freedom $r$ and $R$, respectively 
$\omega_{int}$ and $\omega_{ext}$, were calculated in ref.\cite{quem}, 
together with the ratios (i) and (ii) (see Fig.\ref{raggi}) for different 
$\alpha$.  From the basic observation that a polaron is a bound state of an 
electron plus a phonon cloud, it was pointed out that the frequency of 
vibration of the polaron as a whole is physically limited by the phonon 
frequency ($\omega_{ext}\le \omega_{LO}$).  In other words, at strong 
electron-phonon coupling (or equivalently small $\omega_{LO}$), the lattice 
polarization {\it cannot dynamically follow} the increase in kinetic energy 
induced by the doping, and the quantum fluctuations are transferred to the 
{\it internal} degree of freedom, breaking the polaron apart.  For that 
reason, the polarons are expected to {\it dissociate} at the transition 
for sufficiently high $\alpha$ ($\alpha > \alpha^*
\simeq 7.5$), rather than {\it melting} to a polaron liquid.

We have also calculated the same quantities for an anisotropic polaron.  By 
anisotropic, we mean that the band mass of the electron is different in the 
$xy$ plane than in the $z$ direction, by means of the phenomenological 
parameter $b=m_{xy}/m_z$.  However, the polarization is assumed to preserve 
its three-dimensional character. This is the case for cuprates, as we 
have already dicussed\cite{quem}, on the basis of Chen {\it et al.} 
experiments\cite{chen}. When $b=1$, we recover the usual 3D 
polaron, while for $b=0$, we have some kind of 2D polaron (the electron 
motion is restrained to the $xy$ plane).  Table 1 summarizes the numerical 
results obtained for two different band masses $m^*=m_e$ and $m^*=2m_e$, for 
both $b=1$ and $b=0$.

\vspace{4mm}
\begin{tabular}{l|ccc}
\hline \hline
{} & $\alpha^*$ & $n_c (m^*=m_e)$ & $n_c (m^*=2m_e)$ \\
\hline 
isotropic   $(b=1)$ & $7.5$ & $6 \cdot 10^{18}$ & $ 5 \cdot 10^{19}$\\
(fully) anisotropic $(b=0)$ & $4.4$ & $8 \cdot 10^{19}$ & $ 6 \cdot 10^{20}$\\
\hline \hline
\end{tabular}

\vspace{2mm}
\centerline{\small
Table I. Critical density $n_c (cm^{-3})$ and coupling $\alpha^*$ (see text)}

The important qualitative fact that we can conclude from the numerical 
data is the following: what happens for $\alpha > 7$ in 3D, 
will happen for lower couplings  ($\alpha >4$) in the anisotropic case. In other words, the 
polaron dissociation may be a physical possibility in the cuprates for 
describing the IMT mechanism, at view of the fairly high value of the polar
electron-phonon coupling (Chen {\it et al.}\cite{chen} estimated $\alpha \approx 
5$ for $La_2CuO_4$). 

\pagebreak

\section{Effect of the dipolar interactions: Bagchi's instability}

Let us now concentrate on the melting in the 3D case, and restrain 
ourselves to the limit $\alpha \rightarrow \infty$ ($\omega_{LO} 
\rightarrow 0$).  In this regime, the lattice polarization around each 
electron is frozen, and we can easily include the dipolar interactions that 
were previously neglected.  The electrons vibrate in a potential-well due 
to the static polarization, the jellium sphere and the electrons on the 
other sites.  Assuming that these are localized on the sites of a Bravais 
lattice, the crystal of polarons can be viewed as a {\it Lorentz lattice} 
of dipoles.  Each individual dipole vibrates at the frequency 
$\omega_{int}$ above mentioned.  Bagchi\cite{bagchi} has already calculated 
(classically) the phonon dispersion and the dielectric properties of such a 
Lorentz lattice.  It can be shown\cite{bagchi} that the vibrational 
spectrum $\Omega \left( {\bf k},\lambda \right)$ of such a lattice (${\bf 
k}$ is the wave vector and $\lambda =1,2,3$ is the phonon branch index) 
satisfies the following relations:
\begin{equation} {\Omega }^{2}\left({{\bf k},\lambda }\right) =\ {\omega
}^{2}\left({{\bf k},\lambda}\right) +\ {\omega }_{pol}^{2}, 
\end{equation}
where we have defined ${\omega }_{pol}^{2} = {\omega }_{int}^{2} - {{\omega 
}_{p}^{2}} /{3{\varepsilon }_{\infty }}$, $\omega_p^2=4 \pi n e^2/m^*$ 
being the usual plasma frequency (defined without any dielectric constant).  
$\omega \left( {\bf k}, \lambda \right)$ in (2) is nothing but the phonon 
spectrum of a usual WC.  It satifies the Kohn sum rule\cite{bagchi}: 
$\sum_{\lambda=1}^{3} \omega \left( {\bf k}, \lambda \right)^2 = \omega_p^2 
/ \varepsilon_{\infty}$.  At long wavelengths, there is one 
optical mode ($\omega \left( 0, opt.  \right)= \omega_p / 
\sqrt{\varepsilon_{\infty}}$) and two acoustical modes.  From these properties of 
$\omega \left( {\bf k}, \lambda \right)$ and using (2), it comes out 
immediately that {\it the long wavelength transverse modes of the polaron 
crystal become unstable} when $\omega_{pol}^2 < 0$, i.e.  when 
$\omega_{int} < \omega_p / \sqrt{3 \varepsilon_{\infty}}$.  The critical 
density for which $\omega_{pol}(n_c)=0$, based upon the calculation of 
$\omega_{int}$ in section 2, is $n_c^{Bag.} \approx 5 \cdot 10^{20} cm^{-3}$ for 
$m^*=2m_e$, $\varepsilon_s=30$ and $\varepsilon_{\infty}=5$, which is 
slightly higher than the value obtained through the Lindemann criterion.

The (longitudinal) dielectric constant 
$\varepsilon \left( {\bf k}, \omega \right)$
of the Lorentz lattice has also been calculated by Bagchi.
\begin{equation}
{\frac{1}{\varepsilon \left({\bf k,\mit\omega }\right)}} =
{\frac{1}{\varepsilon_{\infty}}}{\left[1 - {\frac{{\omega
}_{p}^{2}/ {\varepsilon_{\infty}}}{{k}^{2}}}\sum\limits_{\lambda }^{} {\frac{{\left({\bf k\mit\ \cdot
{{\bf e}}_{\bf k,\mit\lambda }}\right)}^{2}}{\left[{{\omega }^{2}\left({\bf
k,\mit\lambda }\right) + {\omega }_{pol}^{2} - {\omega }^{2}}\right]}}\right]}
\end{equation}
where ${{\bf e}}_{\bf k,\mit\lambda }$ is the polarization wavevector of 
the phonon mode ${\bf k},\lambda$.  It can be easily evaluated both in the 
static and in the long wavelength limit. In the latter case, we have:
\begin{equation}
{\varepsilon \left( {0,\mit\omega } \right) }
={\varepsilon_{\infty}}{\left[ 1 - {\frac {\omega_p^2 /{\varepsilon_{\infty}}}
{\omega^2-\omega_{pol}^2} }\right]} ,
\end{equation}
which is sketched in Fig.2.a.  The pole in the dielectric function, located 
at $\omega_{pol}$ (i.e.  the polaron peak in the optical conductivity), is 
shifted to lower energies as the density increases, and tends to zero on 
approaching the IMT, as shown in Fig.2.b.  Strictly speaking, since 
$\omega_{LO}$ is never zero, but assumed to be small, $\omega_{pol}$ should 
rather saturate at higher density.  The corresponding behavior of the 
optical conductivity $\sigma(\omega)=Im \lbrack \varepsilon(0,\omega) 
\rbrack \omega/4\pi$ and energy loss function $Im \lbrack 
-1/\varepsilon(0,\omega)\rbrack$ at $T=0$ is sketched in figure 3.  The 
polaronic peak shift seems to have been recently observed\cite{calvani} in 
the cuprates.  Even though our calculations are strictly valid only in the 
limit ${\omega_{LO}} \rightarrow 0$, one can however argue that for $\alpha 
> \alpha^*$ (above mentioned), the results - in particular the existence of 
Bagchi's instability - remain qualitatively correct, since in that case the 
polarization cannot easily follow the electron vibrations.  New kinds of 
collective modes\cite{ashcroft} corresponding to the mixture of 
polarization waves with electron vibrations can also be expected, as was 
pointed out by Bozovic\cite{bozovic}, which should have an optical 
signature.
%
\begin{figure}[t]
\epsfxsize=11cm 
\centerline{\epsffile{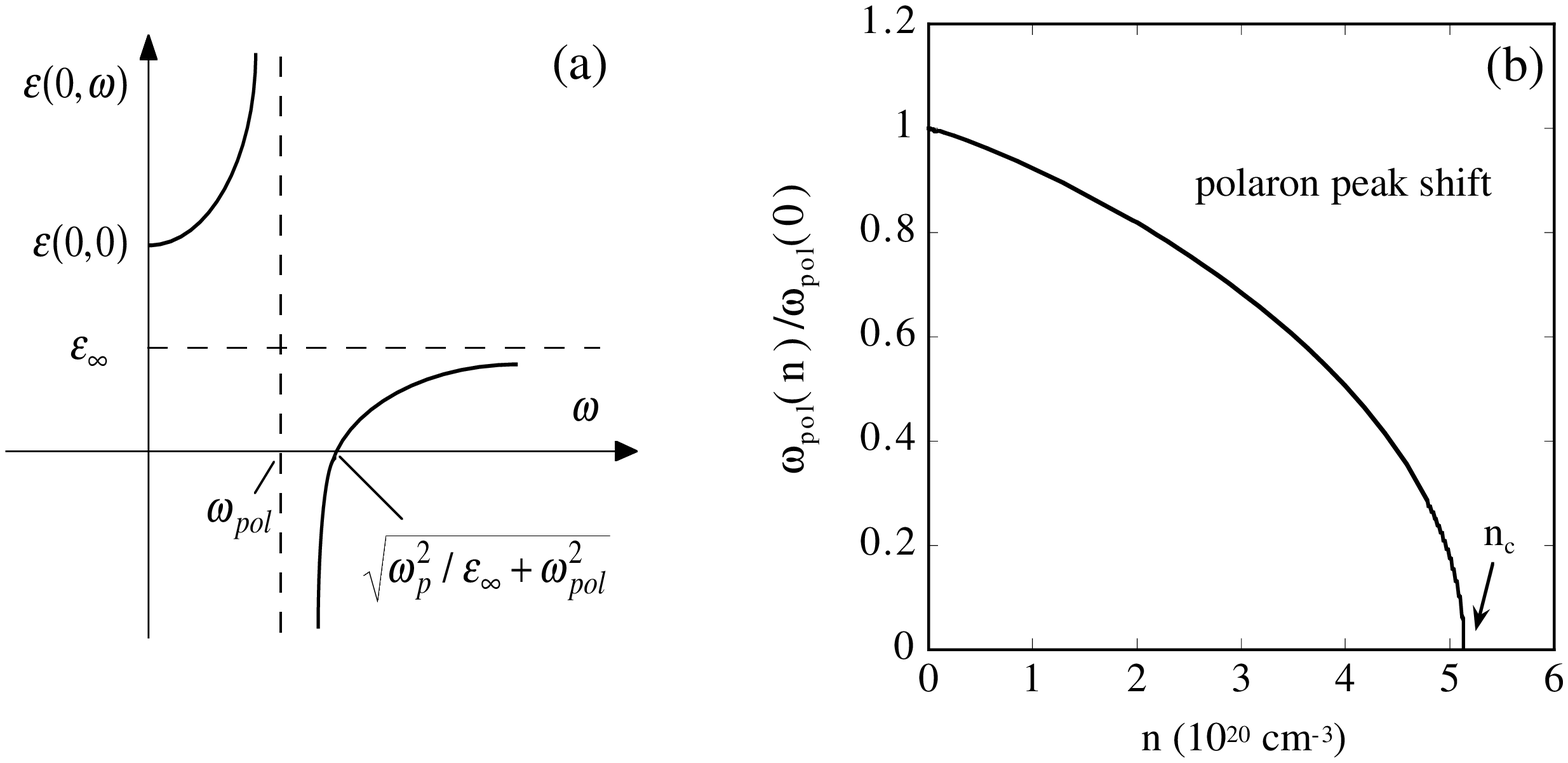}}
\caption{(a) sketch diagram of the long wavelength dielectric function, (b) shift of
the polaron peak and crystal instability in the static limit ($\omega_{LO}=0$), 
for $m^*=2m_e$.} 
 \label{dielectric} \end{figure} 
%
\begin{figure}[t]
\epsfxsize=11cm 
\centerline{\epsffile{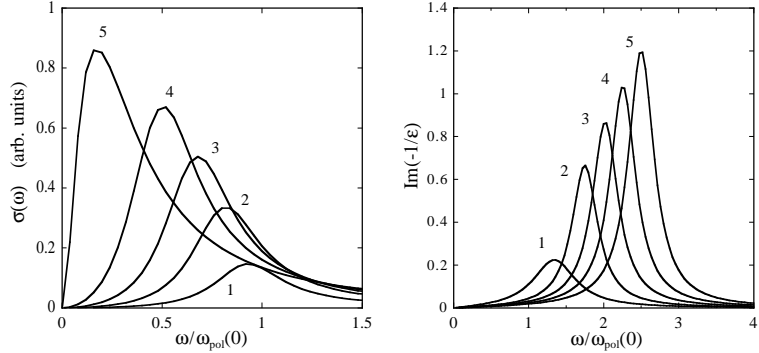}}
\caption{(a) optical conductivity and (b) EELS
spectrum, as calculated by introducing an arbitrary Lorentzian broadening in eq.(4).
Labels 1--5 stand for the density in units of $10^{20} cm^{-3}$. Physical parameters
are the same as in Fig.1, and $n_c\approx 5 \cdot 10^{20} cm^{-3} $.}
\label{spectra} 
\end{figure}
 
As mentioned in the introduction, the static dielectric constant also 
exhibits interesting properties in this limit.  From (3), it can be seen 
that at the critical density ($\omega_{pol}=0$), owing to the sum rule, the 
phonon dispersion obeys $\omega \left({\bf k}, \lambda \right) < \omega_p / 
\sqrt{ \varepsilon_{\infty}}$.  Therefore, we necessarily have $\varepsilon 
\left( {\bf k}, 0 \right) <0$ for any ${ k>0}$.  This indicates the 
possibility of having a net attraction between free carriers coexisting 
with localized polarons beyond the transition.  This possibility opens up 
very interesting questions for both a superconducting scenario, and unusual 
properties of the normal phase beyond the IMT.  Pseudo-separation between 
localized (heavy) polarons and free (light) electrons has already been 
mentioned\cite{quem,remova,bianconi,muller} both theoretically and 
experimentally.  We believe that Bagchi's instability can be one of the 
keys of this pseudo-separation.  Evidently, at the critical density, some 
polarons must be destroyed.  However, it is not necessary to destroy all 
the polaron lattice, and we just need to fulfill the stability condition on 
the vibrational modes $\Omega^2 \left(k,trans.  \right) \ge 0$ (i.e.  
$n_{polarons} \leq n_c^{Bag.}$).  On the other hand, there is some 
energetical argument in favor of a mixed phase of localized polarons with 
free electrons.  The polaron formation energy being negative, it costs 
energy to destroy them.  A compromise might be achieved (which remains to 
be proved) by the coexistence of a polaron lattice with density 
$n_c^{Bag.}$, and quasi-free electrons, provided the mean distance between 
free electrons (corresponding roughly to the screening length) is larger 
than the distance between localized polarons.  However, above a certain 
density (say $n_{free} \approx n_c^{Bag.}$), all the polarons must 
disappear due to a global screening effect.  This last situation could 
correspond to the overdoped region in the cuprates, while the coexistence 
between localized polaron and free electrons, to the underdoped region.

\section{Conclusion}

The studies proposed in this paper (and previous references) are currently 
being extended in several directions, essentially at finite temperature and 
finite phonon frequency, including the dipolar interactions\cite{fratini}.  
The possibility of a mixed state of crystallized polarons coexisting with 
quasi-free carriers is important and must be clarified.  Another problem 
concerns the possibility of having a crystal of bipolarons instead of 
polarons.  This is based on the recent results of Moulopoulos and 
Ashcroft\cite{moulopoulos} for the ordinary WC, where they have proved that at 
intermediate densities, a paired electron crystal is more stable than the 
single electron crystal.  In the polaronic case, this is still an open 
question.  Nevertheless, it would not change the basic result of our studies: at 
strong coupling, the IMT is driven by the dissociation of some localized 
(bi)polarons and cannot yield a liquid state of (bi)polarons.

In conclusion, let us underline that what we have called 
the Bagchi instability, can also be supported by the existence of localized 
dipoles which have nothing to do with polarons.  This remark comes in mind 
if one considers another family of doped oxides, the bismuthates 
($K_xBa_{1-x}BiO_3$), where it has been recently suggested that dipolar 
interactions could be of importance\cite{orlov}.

Part of this work has been presented at the International 
Conference STRIPES in Rome (2-4 june 1998), and took
its origin from very long discussions that one of us (P.Q.) had with A. Bianconi 
during the year 1993, on the possible crystallization of polarons in the
cuprates. This work received financial support from the European
Commission (contract no. ERBFMBICT 961230).

\nonumsection{References}

\end{document}
 
\nonumsection{}

\vspace{-2.5cm}
\begin{figure}[htbp]
\epsfxsize=12cm 
\centerline{\epsffile{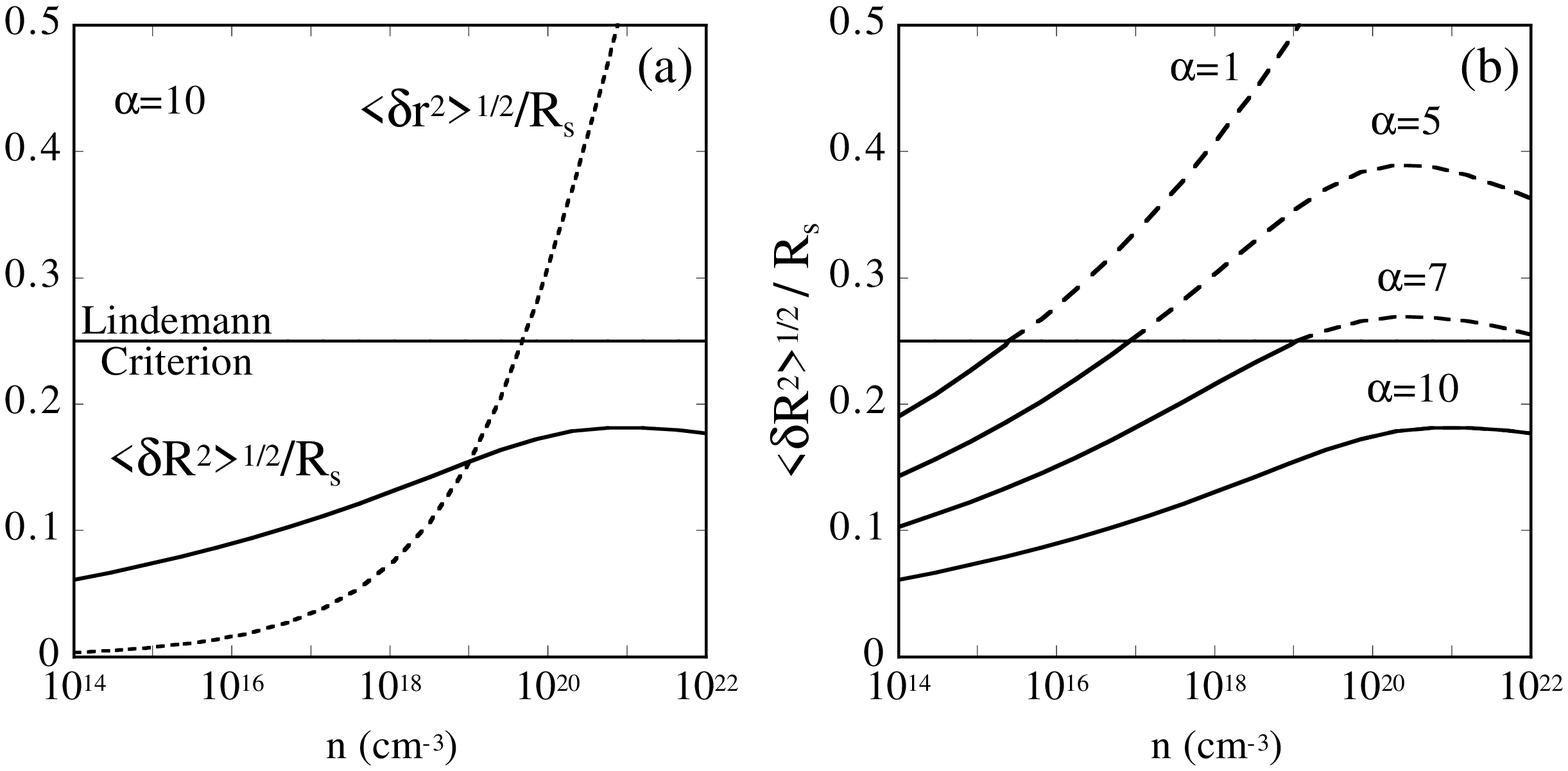}}
\caption{(a)  The fluctuation of the polaron  $\langle \delta R^2
\rangle^{1/2}/R_s$ (bold line) and of its internal degree of freedom $\langle \delta
r^2 \rangle^{1/2}/R_s$ (dotted line) versus $n$ at $\alpha=10$.  (b) The polaron
fluctuation $\langle \delta R^2 \rangle^{1/2}/R_s$ versus $n$ for different $\alpha$.
Physical parameters are $\epsilon_\infty=5$, $\epsilon_s=30$ and $m^*=2m_e$}
\label{raggi} \end{figure}

\vspace{-2.5cm}
\begin{figures}[htbp]
\epsfxsize=11cm 
\centerline{\epsffile{spettri.eps}}
\caption{(a) optical conductivity and (b) EELS
spectrum, as calculated by introducing an arbitrary Lorentzian broadening in eq.(4).
Labels 1--5 stand for the density in units of $10^{20} cm^{-3}$. Physical parameters
are the same as in Fig.1, and $n_c\approx 5 \cdot 10^{20} cm^{-3} $.}
 \label{spectra} 
\end{figure}

\end{}

\end{document}